# Net Proton Flow and Nuclear Transparency Effects at RHIC : A Multiple Fireball Model approach

Saeed Uddin[!], Majhar Ali, Jan Shabir[ŧ]

*Department of Physics, Jamia Millia Islamia, New Delhi-110025*

M. Farooq Mir

*Department of Physics, University of Kashmir, Srinagar*

Recently Beccattini and Cleymamns have proposed a model to understand the $p - \bar{p}$ data obtained at RHIC and SPS. We have shown that this model has a much greater applicability and can be used to describe the rapidity spectra of protons and antiprotons separately measured at the highest energy of Relativistic Heavy Ion Collider, $\sqrt{S_{NN}}$ = 200 GeV by BRAHMS Collaboration. We have also determined that the contribution of the decay products of the heavier resonances like Δ, Λ etc. actually dominate over the protons (antiprotons) of pure thermal origin. We have also imposed the criteria of exact strangeness conservation in each fireball separately. It is shown that it is possible to explain not only the ($p - \bar{p}$ ) data but the complete set of data viz. the rapidity distributions of protons, antiprotons and the ratio $\bar{p}/p$ simultaneously quite well with a single set of model parameters and a single value of the temperature parameter T chosen for all the fireballs. We have also fitted the Kaon rapidity spectra obtained by BRAHMS collaboration in the same experiment. This is unlike the previous analyses wherein a varying temperature has been assumed for the fireballs formed along the rapidity axis. The chemical potentials are however assumed to be dependent on the fireball rapidity $y_{FB}$.

[!] E-mail : saeed_jmi@yahoo.co.in
[ŧ] On deputation from Department of Physics, Amar Singh College, Srinagar.

==================================================================================

The yields of baryons and antibaryons are an important indicator of the multi-particle production phenomenon in the ultra-relativistic nucleus-nucleus collisions. Great amount of experimental data have been obtained in such experiments ranging from the AGS energies to the RHIC. The study of ultra-relativistic nuclear collisions allows us to learn how baryon numbers, initially carried by the nucleons, are distributed in the final state [1]. It is possible to obtain important information about the energy loss of the colliding nuclei by analyzing the rapidity dependence of the p and $\bar{p}$ production. The measurement of the *net* proton flow (i.e. $p - \bar{p}$ ) in such experiments can throw light on the collision scenario, viz. the extent of nuclear stopping/transparency, formation of the fireball and the degree of thermo-chemical equilibration of heavier resonances (Δ etc.) which can be estimated by their contribution to lighter hadrons, say, $p$ and $\bar{p}$ via their decay. The net proton flow at the AGS energies peaks at midrapidity, while at the top SPS energy ($\sqrt{S_{NN}}$ ≈17.3 GeV) the distribution develops a minimum at midrapidity. The SPS data at different energies (20, 30, 40, 80, 158 GeV/A) show [1] that at midrapidity the $p$ yield decreases gradually with increasing energy in contrast to the rapidly rising $\bar{p}$ yield. This implies that at SPS energies nuclear collisions start exhibiting some transparency. Recently a new property has emerged namely the extended longitudinal scaling in the rapidity distributions [2 - 5]. This has been observed in pp collisions as well as ultra-relativistic collisions at the highest RHIC energies [5, 6]. Data from the BRAHMS collaboration [7] also show that the antiproton to proton ratio shows a maximum at mid rapidity and gradually decreases towards larger rapidities, whereas the net proton flow shows a broad minimum, spanning about ± 1 unit around mid-rapidity region of dN/dy spectra. It was therefore conjectured [7] that at RHIC energies the collisions are (atleast partially) transparent. Though the midrapidity region at RHIC is not yet totally baryon free however a transition from a baryon dominated system at lower energies to an *almost* baryon free system in the midrapidity at RHIC can be observed. An interesting analysis by Stiles and Murray [2,8] shows that the data obtained by the BRAHMS collaboration at 200 GeV has a clear dependence of the baryon chemical potential on rapidity due to the changing $\bar{p}/p$ ratio. Biedroń and Broniowski [2,9] have done an analysis of rapidity dependence of the $\bar{p}/p$, $K^+/K^-$, $\pi^+/\pi^-$ ratios based on a single freeze-out model of relativistic nuclear collisions. Fu-Hu Liu *et al.* have [10] recently attempted to describe the transverse momentum spectrum and rapidity distribution of net protons produced in high energy nuclear collisions by using a new approach viz. a two cylinder model [11 - 13]. In order to describe the rapidity distribution of the produced hadrons in ultrarelativistic nuclear collisions the statistical thermal model has been extended to allow for the chemical potential and temperature to become *rapidity dependent* [5, 9, 14 - 16]. Recently Becattini *et al.* and Cleymans *et al.* [2, 5, 14 - 16] have attempted to describe the net proton flow data obtained at RHIC in Au – Au collisions at 200 GeV. They have applied the thermal model in a very interesting way where they consider the rapidity axis to be populated with fireballs moving along it with increasing rapidity, $y_{FB}$. The emitted particles leave the fireballs at freeze-out following a thermal distribution. The rapidity distribution of any given particle specie $j$ can then be written as:

$$\frac{dN^j(y)}{dy} = \int_{-\infty}^{+\infty} \rho(y_{FB}) \frac{dN_1^j(y - y_{FB})}{dy} dy_{FB} \qquad (1)$$

where y is the particle's rapidity. The distribution $\frac{dN^j(y)}{dy}$ represents total contribution of all the fireballs to the $j^{th}$ hadron specie's rapidity spectra. This also includes the contribution of the decay products of heavy resonances. The contribution of the respective fireballs follows a Gaussian distribution centred at zero fireball rapidity ($y_{FB}$ = 0) :

$$\rho(y_{FB}) = \frac{1}{\sqrt{2\pi}\sigma} \exp\left(\frac{-y_{FB}^2}{2\sigma^2}\right) \qquad (2)$$

The value of σ determines the width of the distribution. The data essentially require a superposition of fireball contributions along the rapidity axis where the baryon chemical potential ($\mu_B$) of the successive fireballs is dependent on the fireball rapidity ($y_{FB}$). A quadratic type dependence is considered [2, 4] :

$$\mu_B = a + b\, y_{FB}^2 \qquad (3)$$

In the recent works [5, 17] it has been further assumed that the temperature of the successive fireballs along the rapidity axis *decreases* (as the baryon chemical potential increases) according to a chosen parameterization :

$$T = 0.166 - 0.139\,\mu_B^2 - 0.053\,\mu_B^4 \qquad (4)$$

where the units are in GeV. Here the temperature of the mid-rapidity fireball ($y_{FB} \sim 0$) is fixed at ~166 MeV.

In the work of Becattini and Cleymans [2, 4] a good fit to the net proton flow as measured at the highest RHIC energy by the BRAHMS collaboration has been obtained. The values of the model parameters fitted by them at the highest RHIC energy are, a = 23.8 MeV, b = 11.2 MeV and σ = 2.183. The temperature T varies according to the parameterization (4). However, there is no fit provided for the $\bar{p}/p$ data or the proton and antiproton data independently. Moreover the strange sector data as measured by the BRAHMS collaboration in the same experiment for the same experimental conditions are not described at all by these authors.
Therefore in this letter we have attempted to use the above discussed multiple fireball model to explain the proton flow, anti proton flow, the net proton flow (i.e. $p - \bar{p}$ ), the ratio $\bar{p}/p$ , the Kaon and the AntiKaon flows ***simultaneously***. It is shown that this can be achieved with a single set of model parameters (viz. a, b, σ) which also includes a single value of the temperature parameter T chosen for all the fireballs. This is unlike the previous analyses wherein a varying temperature has been assumed for the fireballs formed along the rapidity axis. The chemical potentials are however still assumed to be dependent on the fireball rapidity $y_{FB}$, a situation which is unavoidable in the model, due to the nature of the data. Hence we have shown that the model proposed by Beccattini and Cleymans [2,4] has a much greater applicability than previously thought.
In figures 1 and 2 we have shown the experimental dN/dy data (by the red solid boxes) for the protons and antiprotons, respectively, obtained from the top 5% most central collisions at $\sqrt{S_{NN}}$ = 200 GeV in the BRAHMS experiment. The errors are both statistical and systematic. The proton and antiproton *dN/dy* decrease from midrapidity to *y* ~ 3. We have fitted both spectral shapes simultaneously for a = 23.5, b = 15.2, σ = 1.89 and T = 177.0 MeV.

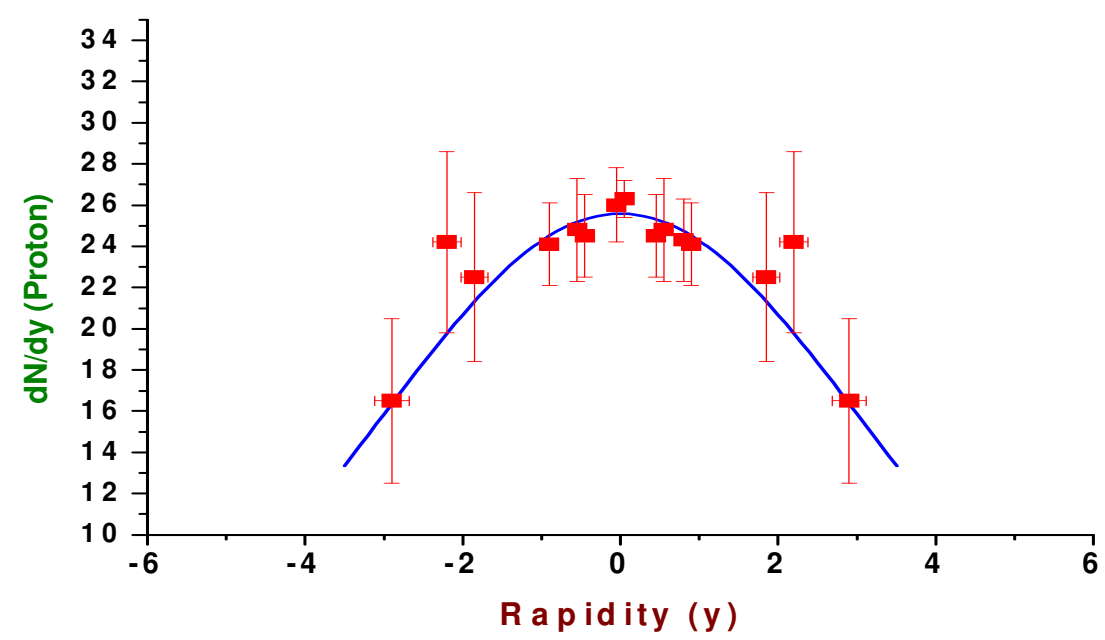

***Fig 1. : Proton rapidity spectra. Theoretical result is shown by the solid curve.***

We find that the theoretical curves fit the data quite well in both the cases. The experimental data have been symmetrized for the negative values of rapidity [7]. The weighted $\chi^2$/DoF for the two fits in figures 1 and 2 are 0.23 and 0.48, respectively.
Moreover the proton spectra in the figure 1 is seen to be slightly broader than the antiproton spectra in figure 2. This according to the present model seems to emerge from the fact that since $\mu_B \sim y_{FB}^2$ and the rapidity axis is assumed to be populated by the fireballs of successively increasing rapidity $y_{FB}$ and hence increasing chemical potentials, consequently the low rapidity (y) baryons (which have a larger population in a baryon rich fireball in thermo-chemical equilibrium) emitted in the forward (backward) direction from a *fast* fireball (i.e. fireballs with large $y_{FB}$), appear with a large value of rapidity (y) in the *rest frame of the colliding nuclei*. In other words as the baryon chemical potentials ($\mu_B$) increase monotonically along the rapidity axis (as ~ $y_{FB}^2$) there is an increase in the *density* of the protons and a simultaneous suppression in the *density* of antiprotons.

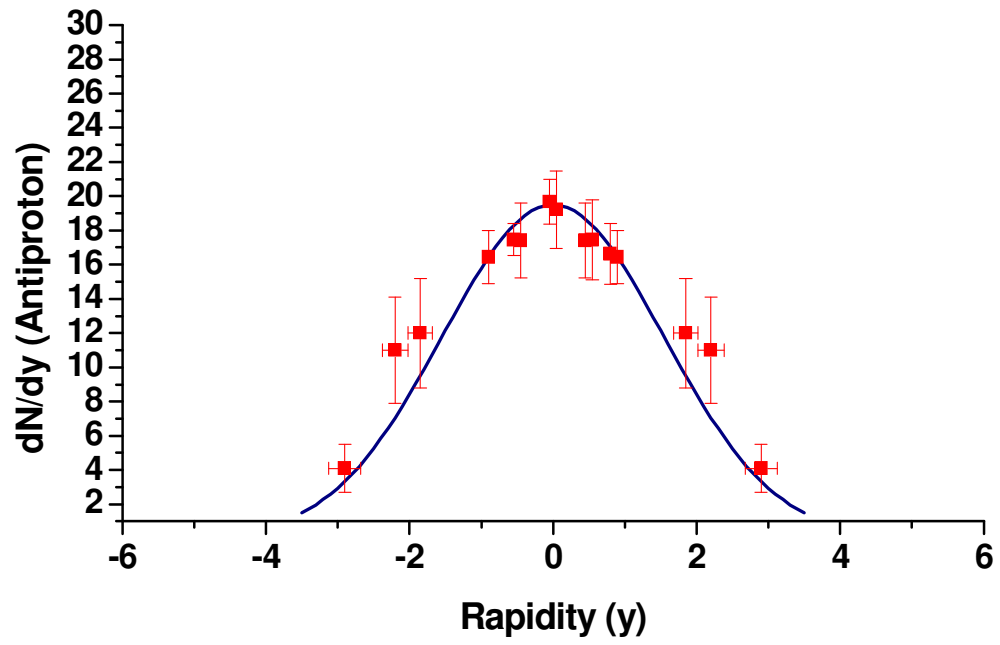

***Fig 2. : Antiproton rapidity spectra. Theoretical result is shown by the solid curve***

Figure 3 shows the rapidity spectra dN/dy for the net proton flow. The theoretical curve which fits the data for the *same* values of the model parameter mentioned above is also shown. There is somewhat a broad minimum around the midrapidity region. However the situation is not as was expected widely [18] that the rapidity distribution of baryons produced in the ultra-relativistic nuclear collisions will exhibit a very flat minimum, centred at midrapidity. So far it has not been observed, either in the SPS experiments or at the RHIC. The $x^2$/DoF is 0.56 in this case.

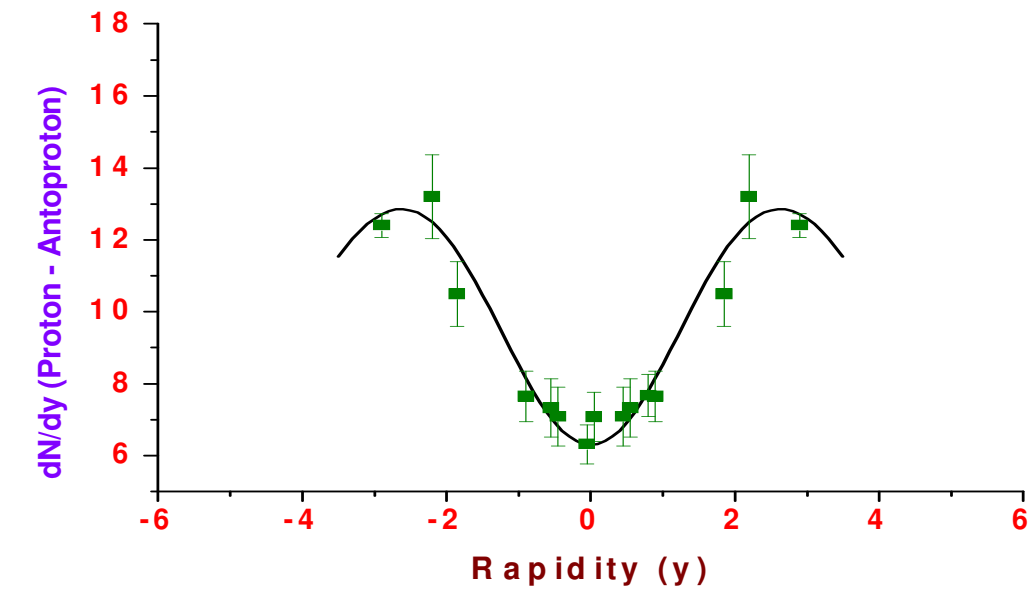

***Figure 3 : Rapidity spectra of the net proton flow. The theoretical curve is for the same values of the model parameters as in figures 1 and 2.***

In figure 4 we have shown the rapidity spectra of the $\bar{p}/p$ ratio. The ratio has somewhat a broad maximum (~ 0.75) in the midrapidity region which then decreases to about 25% at around y ~ 3. The theoretical curve shown is again for the *same* values of the parameters as mentioned above.

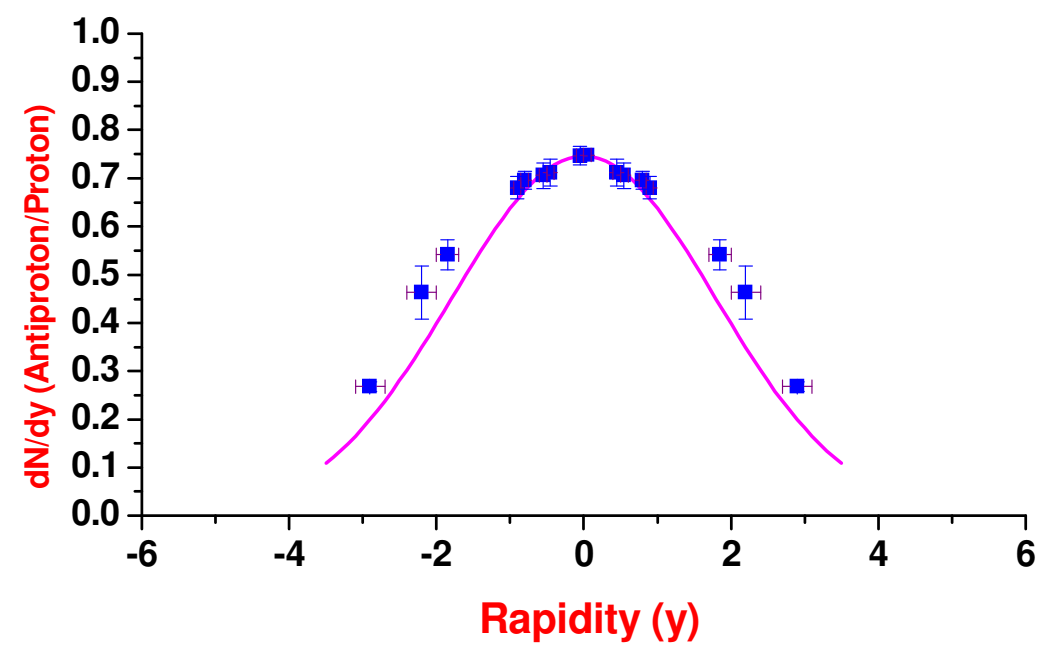

***Figure 4 : Rapidity spectra of antiproton/proton ratio. The calculated rapidity distribution is shown by the curve.***

We find that the curve provides a reasonably good fit to the experimental data. The $x^2$/DoF is 0.17 in this case.

In the analysis we have also included the contribution of the decay protons and antiprotons (from the Δ, Λ etc.). It is interesting to find that the contributions of the decaying hadronic resonances actually dominate over the protons and antiprotons of pure thermal origin (i.e. those which are not the decay products).

We have also analyzed the strange meson data as measured by the BRAHMS collaboration in the top 5% most central collisions at 200 GeV/A in the same Au + Au collision experiments. We have found that in the above model it is possible to account for the rapidity distribution of (Anti) Kaons also.

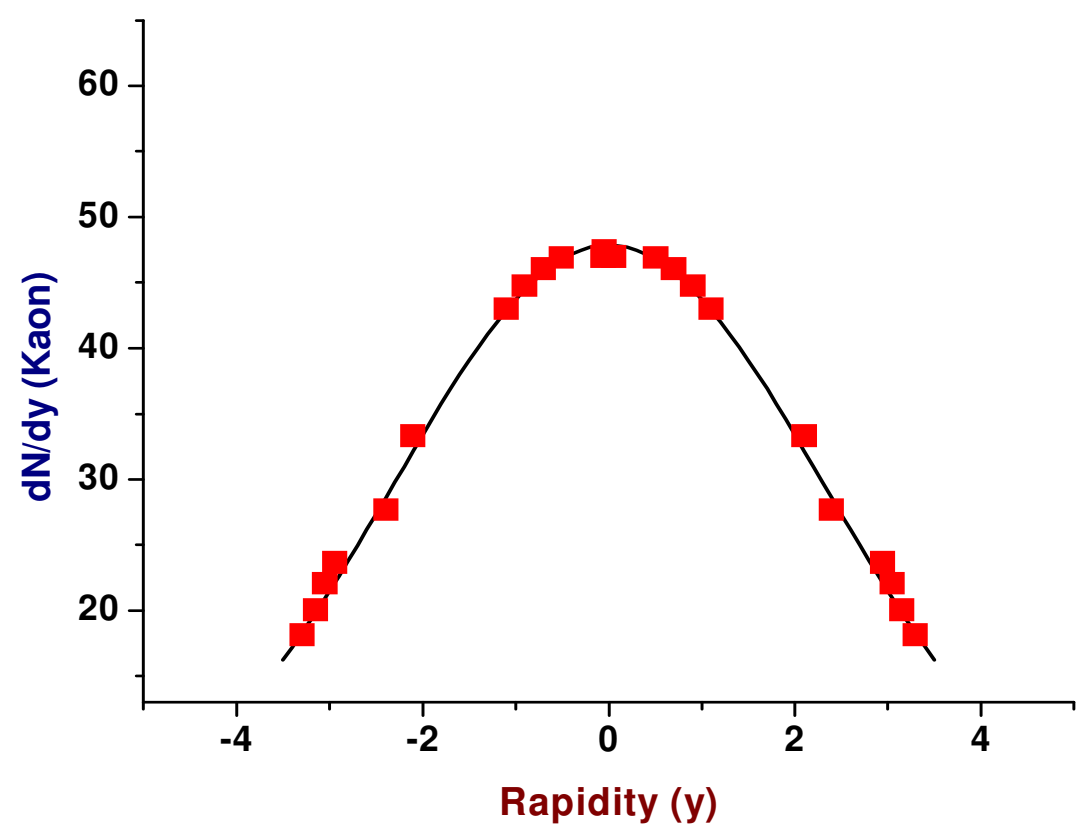

***Figure 5 : Rapidity spectra of Kaon flow. The theoretical curve which fits the data is for the same values of the model parameters as used for the theoretical curves in figures 1 and 2. The error bars are small and contained within the solid red data point boxes.***

In figure 5 we have shown the rapidity spectra of Kaon flow. The theoretical curve which fits the data is for the *same* values of the model parameters as used for the theoretical curves in figures 1 and 2. We find that the theoretical curve provides a good fit to the data.

In figure 6 we have shown the rapidity spectra of AntoKaons. The theoretical curve which fits the data is again for the same values of the model parameters used in figures 1 and 2.

We find that the Kaon spectra is broader than the AntiKaon spectra as also noticed for the proton spectra which is broader than the antiproton spectra. This is again due to the increasing chemical potential of the successive fireballs.

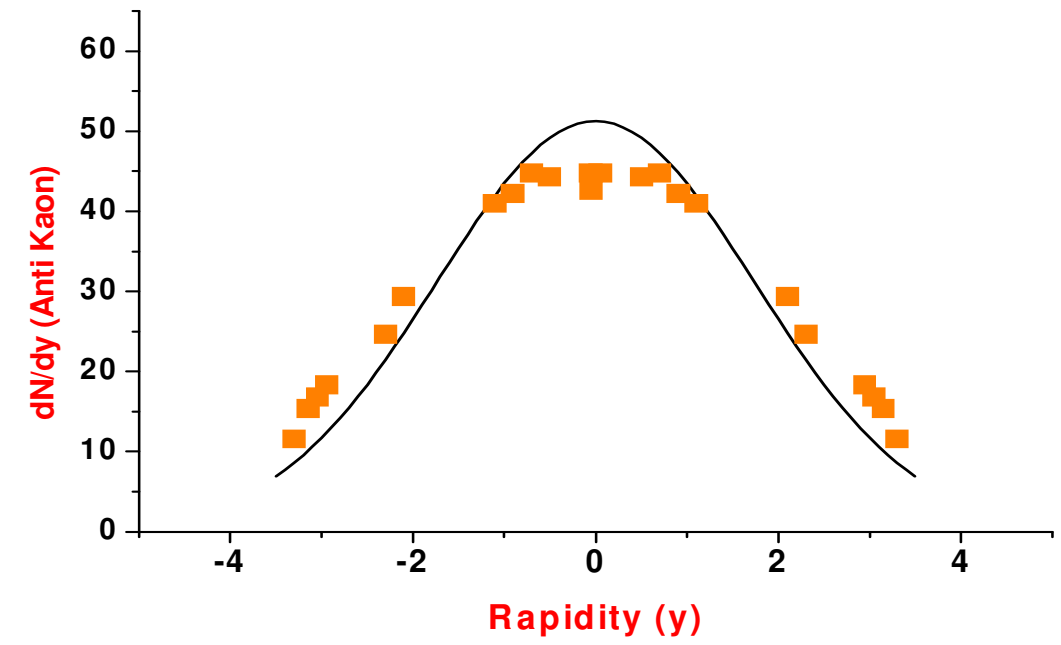

***Figure 6 : Rapidity spectra of AntiKaon flow. The theoretical curve which fits the data is for the same values of the model parameters as used for the theoretical curves in figures 1 and 2. The error bars are small and contained within the solid red data point boxes.***

The RHIC data also show that the $K^*/K$ ratio in the Au + Au collisions is about 22.8%. In the present model calculation we find that the theoretically calculated value of $K^*/K$ ratio is approximately 22.5%. Hence approximately 22.5% Kaons are the decay products of the heavier $K^*$ resonance. Similarly the analysis also shows that nearly 70% of the protons are trapped inside the heavier resonances like $\Delta$, $\Lambda$, $\Sigma$ etc.

In our analysis we have applied the criteria of exact strangeness conservation. It is done in a way such that the net strangeness is zero not only on the overall basis but also in every fireball separately. This is essential because as the rapidity of the fireballs formed increases along the rapidity axis the baryon chemical potential ($\mu_B$) increases. Hence the required value of the strange chemical potential ($\mu_S$) varies accordingly for each fireball for a given value of temperature T (= 177.0 MeV here). Consequently the values of the strange chemical potential ($\mu_S$) will vary with $y_{FB}$.

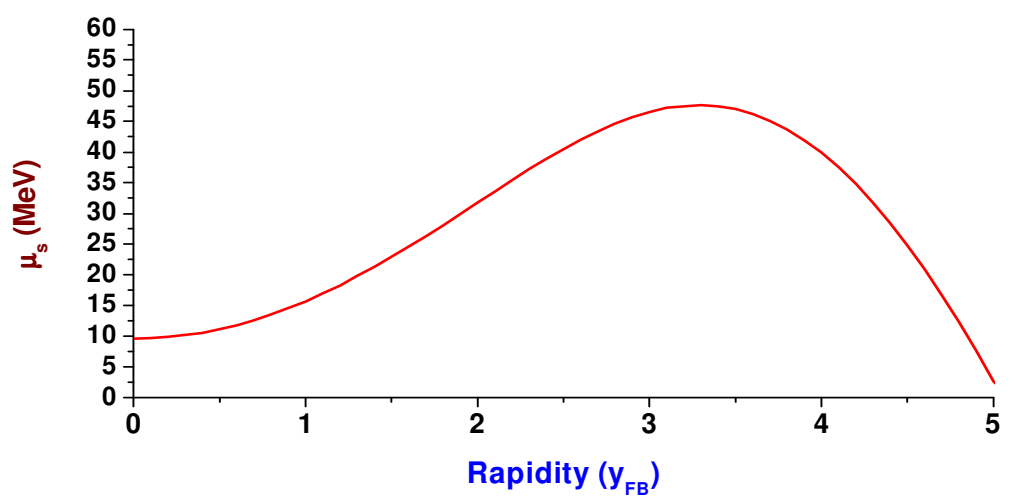

***Figure 7 : The variation of the $\mu_S$ with $y_{FB}$ for T = 177.0 MeV.***

In figure 7 we have shown this by plotting the variation of the $\mu_S$ with $y_{FB}$. It is seen to first rise smoothly with $y_{FB}$ reaching a maximum value of about 50 MeV at around $y_{FB}$ ~ 3.5 and then drops rapidly to very small values as $y_{FB}$ ~ 5.

In summary, we use an extended thermal model proposed earlier where formation of several fireballs moving with increasing rapidity ($y_{FB}$) along the rapidity axis is assumed. A Gaussian profile in $y_{FB}$ is used to provide a weight factor to estimate their contribution to the emitted hadrons population. A quadratic profile in $y_{FB}$ is used to fix the baryon chemical potentials of these fireballs. We find that it is possible to explain not only the net proton flow but also the proton and antiproton rapidity spectra separately as well as the $\bar{p}/p$ ratio simultaneously. Furthermore it is interesting to find that the model is also very successful in explaining the strange meson data quite well measured in the same experiment by the BRAHMS collaboration without requiring any change in the values of the model parameters. Hence this is achieved by using a *single* set of the model parameters. This is unlike the previous analyses of the RHIC data where theoretical fits to the proton, antiproton, $\bar{p}/p$ ratio, Kaon and the AntiKaon spectra were not shown and the temperature T of the individual fireballs was also assumed to vary continuously with $y_{FB}$.

***Authors are grateful to Professors J. Cleymans and F. Becattini for making the experimental data available to us. Jan Shabir is grateful to University Grants Commission, New Delhi, for the financial assistance during the period of deputation. Majhar Ali is thankful to Jamia Millia Islamia for providing scholarship.***

References:

[1] E. Kornas *et al.* NA 49 Collaboration, Eur. Phys. J. ***C49*** (2007) 293.

[2] F. Becattini *et al.* arXiv : 0709.2599v1 [hep-ph].

[3] Fu-Hu Liu *et al.*, Europhysics Letters **81** (2008) 22001.

[4] J. Cleymans, J. Phys ***G35*** (2008) 1.

[5] J. Cleymans *et al.* arXiv : 0712.2463v4 [hep-ph] 2008.

[6] G.J. Alner *et al.* Z. Phys. ***C33*** (1986) 1.

[7] I.G. Bearden *et al.*, BRAHMS Collaboration, Phys. Rev. Lett. **93** (2004) 102301.

[8] L.A. Stiles and M. Murray, nucl-ex/0601039.

[9] B. Biedroń and W. Broniowski, Phys. Rev ***C75*** (2007) 054905.

[10] Fu-Hu Liu *et al.*, Europhys. Lett. **81** (2008) 22001.

[11] Fu-Hu Liu *et al.*, Phys. Rev ***C69*** (2004) 034905.

[12] Fu-Hu Liu, Phys. Rev. ***C66*** (2002) 047902.

[13] Fu-Hu Liu, Phys. Lett. **B583** (2004) 68.

[14] F. Becattini and J. Cleymans, J. Phys. ***G34*** (2007) S959.

[15] F. Becattini *et al.*, Proceedings of Science, CPOD07 (2007) 012,

[16] W. Broniowski and B. Biedroń J. Phys. ***G35*** (2008)044018, arXiv:0709.0126 [nucl-th].

[17] J. Cleymans *et al.* Phys. Rev. ***C75*** (2006) 034905.

[18] J.D. Bjorken, Phys. Rev. **D27** (1983) 140.